\documentclass[12pt]{article}

\usepackage{a4wide}

\usepackage{amsmath,amssymb,amsfonts} 

\usepackage{extarrows}
\usepackage{undertilde}

\usepackage{mathrsfs,latexsym}

\newcommand{\Ett}{{\tt E}}

\newcommand{\CC}{{\mathbb C}}

\newcommand{\RR}{{\mathbb R}}

\newcommand{\NN}{{\mathbb N}}

\newcommand{\ZZ}{{\mathbb Z}}

\newcommand{\CoinfM}{C_0^\infty(M)}

\newcommand{\CoinX}[1]{C_0^\infty({#1})}

\newtheorem{Thm}{Theorem}[section]

\newtheorem{Prop}[Thm]{Proposition}

\numberwithin{equation}{section}

\newcommand{\Cc}{{\mathcal C}}

\newcommand{\kb}{{\boldsymbol{k}}}

\newcommand{\ip}[2]{{\langle #1\mid #2\rangle}}

\newcommand{\bra}[1]{{\langle #1 \mid}}

\newcommand{\ux}{\underline{x}}

\newcommand{\Mb}{{\boldsymbol{M}}}
\newcommand{\Nb}{{\boldsymbol{N}}}

\newcommand{\Sc}{{\mathcal{S}}}

\newcommand{\Af}{{\mathscr A}}

\newcommand{\Df}{{\mathscr D}}

\newcommand{\Ff}{{\mathscr F}}

\DeclareMathOperator{\sinc}{sinc}

\begin{document}

%

\title{On a Recent Construction of ``Vacuum-like'' Quantum Field States
in Curved Spacetime}
\author{Christopher J Fewster${}^{(1)}$\thanks{\tt chris.fewster@york.ac.uk}~
and Rainer Verch${}^{(2)}$\thanks{\tt verch@itp.uni-leipzig.de}\\[12pt]
\small ${}^{(1)}$ Department of Mathematics,
                 University of York,
                 Heslington,
                 York YO10 5DD, U.K. \\
\small ${}^{(2)}$ Institut f\"ur Theoretische Physik,
Universit\"at Leipzig,
04009 Leipzig, Germany}
\date{\small \today}
\maketitle
\begin{abstract}
Afshordi, Aslanbeigi and Sorkin have recently proposed a construction
of a distinguished ``S-J state'' for scalar field theory in (bounded regions of)
general curved spacetimes. We establish rigorously that the proposal
is well-defined on globally hyperbolic spacetimes or spacetime
regions that can be embedded as relatively compact subsets of other globally
hyperbolic spacetimes, and also show that, whenever the proposal is
well-defined, it yields a pure quasifree state. However, by explicitly
considering portions of ultrastatic spacetimes, we show that the S-J state
is not in general a Hadamard state. In the specific case where the 
Cauchy surface is a round $3$-sphere, we prove that the representation induced by 
the S-J state is generally not unitarily equivalent to that of a Hadamard state, and 
indeed that the representations induced by S-J states on nested regions
of the ultrastatic spacetime also fail to be unitarily equivalent in general.
The implications of these results are discussed. 
\end{abstract}

\section{Introduction}
A striking difference between quantum field theory in flat and curved spacetimes
is the role of the quantum state. In flat spacetime, the Poincar\'e invariant
vacuum state is, in many ways, the bedrock of the theory and its high
degree of symmetry can be used to explain many characteristic features of the
theory, such as the connection between spin and statistics and the existence of a PCT symmetry~\cite{StreaterWightman, Haag}. The situation is very different in curved spacetime, for a generic spacetime has no symmetries at all and there is, consequently, no single state picked out by its symmetry properties. Moreover, as the Fock space built on the Minkowski vacuum vector is intimately related to the description of particle states, the absence of a preferred state in curved spacetime calls into question the very notion of a particle. 

The standard view is that this is no bad thing: from an operational perspective, particles are what particle detectors detect, and the motion of a particle detector critically influences the results it produces, as Unruh showed long ago~\cite{Unruh:1976}. So, even in Minkowski space, it is not clear that  the particle interpretation should be an essential building block of the theory, rather
than an emergent feature in suitable circumstances. The absence of a preferred state can be viewed as an opportunity to refocus attention on the essential content 
of the theory.\footnote{As Fulling~\cite{Fulling:1979} put it: `With the field itself thereby put in its rightful place at the center of the physical interpretation of the theory, the particle concept is reduced to a somewhat arbitrary technical aid'
(although, as he remarks, there can be uses for it). The algebraic approach to quantum field theory~\cite{Haag} takes this a step further, focussing on local observables, with quantum fields as auxiliary objects.} Instead of seeking a single preferred state, the focus has shifted to a preferred class of physically acceptable states, with the class of Hadamard states emerging as the leading contender. The Hadamard states are preferred both for their numerous technical properties and because they permit the computation of expectation values for physical quantities such as 
the stress-energy tensor, and indeed, for the perturbative construction of interacting quantum field models~\cite{BrFr2000,Ho&Wa01,Ho&Wa02}. 

Nonetheless, there are good reasons to seek natural constructions of quantum states in curved spacetime. For one thing, it is notoriously difficult to provide concrete examples of Hadamard states in general spacetimes, though it is known that they are abundant~\cite{FullingNarcowichWald}. For another, it is of clear interest to select, from among the Hadamard states, those that are `as vacuum-like as possible' (or, for other purposes, `as thermal as possible'). In the past, various attempts have been made to define distinguished states, for instance by diagonalising the Hamiltonian, or attempting to determine an instantaneous notion of positive frequency. Such attempts are typically either ill-posed on closer inspection (Hamiltonian diagonalisation) or fail to produce Hadamard states (instantaneous vacua). A number of proposals are critically discussed in~\cite{Fulling:1979}; we also refer to the cautionary comments in~\cite{Kay:2006}. Moreover, we have recently proved a no-go theorem that shows, in essence, that {\em no} reasonable  quantum field theory formulated in all spacetimes can admit a single preferred choice of state in each~\cite{FewVer:dynloc_theory}. We will return to this later.

A novel approach has recently been proposed by Afshordi, Aslanbeigi and Sorkin~\cite{AAS}, inspired by constructions originating in the causal set approach to quantum gravity~\cite{Johnston:2009,Sorkin:2011}. The idea is to use the Pauli--Jordan commutator function (the difference of the advanced and retarded Green functions) to define a state associated with any
bounded globally hyperbolic spacetime region, called the `S-J vacuum' after
the authors of~\cite{Johnston:2009,Sorkin:2011}.  In this paper, we show that the resulting
 state is well-defined and pure in a wide range of circumstances. However, we also show that it
fails to be Hadamard in general. This is done by an explicit computation of the state
on `ultrastatic slab' spacetimes with compact spatial section, on which the S-J prescription is well-defined.
By an ultrastatic slab, we mean a spacetime $I\times\Sigma$ equipped 
with an ultrastatic 
metric, where $I\subset\RR$ is a relatively compact open interval;
 such manifolds may be regarded either as spacetimes in their own right, 
or bounded regions of a larger spacetime. 
In fact,~\cite{AAS} considers quite a similar calculation in a discussion 
of the S-J prescription on 
the static spacetimes, which must be approached via a limit 
over similar spacetime slabs (incidentally, we give a clean treatment 
of that issue in the ultrastatic case). Even more, 
in the case 
where $\Sigma$ is the 3-sphere, we show that generically (i) the 
S-J state fails to be unitarily equivalent to the ultrastatic Hadamard vacuum and
(ii) the S-J states defined with respect to different time intervals $I$ 
fail to be unitarily equivalent. Failure of unitary equivalence in this context
means, in particular, that the said states induce incompatible notions
of ``vacuum'' and ``particles''; therefore, these results cast doubts on the
significance of S-J states as physical states.
Finally, we will discuss how the S-J construction evades the no-go theorem mentioned above, and the implications this has for its interpretation.

\section{The Quantized Scalar Field on Globally Hyperbolic Spacetimes}

We wish to investigate and discuss the proposal of~\cite{AAS} for distinguished states of
the linear scalar field on curved spacetimes in a more general (and more mathematically
oriented) setting and thus we will start by collecting some material on the Klein-Gordon field
in globally hyperbolic spacetimes. In particular, we will summarize some basics on the
CCR algebra associated with the quantized Klein-Gordon field, its (quasifree) states and the
corresponding Hilbert space representations. All this material is standard and has been
documented elsewhere; for our summary, we draw in particular on~\cite{BarGinouxPfaffle,
KayWald:1991,Wald_qft,Verch:1994,ArakiYamagami:1982,ManuVerb:1968}.

\subsection{Globally hyperbolic spacetimes}

A spacetime $\Mb = (M,g_{ab})$ is an $n = 1 +d$-dimensional smooth manifold, connected and orientable,
with a Lorentzian metric $g_{ab}$. The metric signature will be taken as $(+,-, \ldots,-)$.
A spacetime is called {\em globally hyperbolic} if it admits smooth foliations in smooth spacelike Cauchy-surfaces, which are smooth co-dimension $1$ submanifolds of $M$ intersected exactly once by every inextendible causal curve. Globally hyperbolic spacetimes
are time-orientable and we always assume that some time-orientation has been chosen; 
the symbol $\Mb$ collects together the spacetime manifold $M$ together with Lorentzian metric $g_{ab}$,
orientation and time-orientation. For any subset $G \subset M$, one denotes by
$J^\pm(G)$ the causal future(+)/past$(-)$ sets of $G$ in $\Mb$, i.e.\ the sets of all points in
$M$ which can be reached by any future(+)/past$(-)$ directed causal curve in $\Mb$ starting
at some point in $G$. Denoting by $d{\rm vol}_{\Mb}$ the volume form on $M$ induced by $g_{ab}$,
there is a natural scalar product on the space $\CoinfM$ of smooth, compactly supported, 
complex valued functions on $M$, given by
\begin{align}
 \langle f, h \rangle & = \int_M \overline{f} h\,d{\rm vol}_{\Mb}\,, \quad f,h \in \CoinfM\,,
\end{align}
where overlining denotes complex conjugation.
The Hilbert space obtained from $\CoinfM$ and this scalar product is $L^2(M,d{\rm vol}_{\Mb})$,
denoted for short just by $L^2(\Mb)$.

A spacetime $\Mb$ will be said to be {\em ultrastatic} if it takes the form $M = \RR\times\Sigma$
with metric $ds^2 = dt^2 - h_{ij}(\ux)dx^i dx^j$ where the $x^i$ are local coordinates
on $\Sigma$ and $h_{ij}$ is a Riemannian metric on $\Sigma$, independent of the time coordinate $t$. 
If $\Sigma$ is compact (or, more generally, if $h$ defines a complete metric on $\Sigma$) the
spacetime is globally hyperbolic~\cite{Kay1978}; of course we take the time-orientation so that $\partial/\partial t$ is future-pointing. A spacetime of the form $I\times\Sigma$,
where $I\subset\RR$ is a relatively compact open interval, equipped with
an ultrastatic metric will be called an {\em ultrastatic slab}.

\subsection{Klein-Gordon equation}

Let $\Mb = (M,g_{ab})$ be a globally hyperbolic spacetime. Then we refer to
the partial differential operator $\Box + m^2 : C^\infty(M) \to C^\infty(M)$, where
$\Box = \nabla^a \nabla_a$ is the d'Alembertian operator on scalar functions associated
to $\Mb$, and $m \ge 0$ is a fixed constant, as {\em Klein-Gordon operator}, and to
\[
 (\Box + m^2)\varphi = 0
\]
as {\em Klein-Gordon equation} on $\Mb$. For globally hyperbolic $\Mb$, the 
Cauchy-problem for the Klein-Gordon equation is well-posed. Equivalently, there are
unique advanced($-$)/retarded($+$) fundamental solutions $\Ett^\pm$ of the 
Klein-Gordon operator. These are linear maps
\[
  \Ett^\pm : \CoinfM \to C^\infty(M)\,,
\]
which are characterized by being continuous (with respect to appropriate test-function
topologies, see~\cite{BarGinouxPfaffle} for details), and having the following properties for
all $f \in \CoinfM$:
\begin{align*}
 (\Box + m^2) \Ett^\pm f & = f = \Ett^\pm (\Box + m^2) f \,, \\
       {\rm supp}(\Ett^\pm f) & \subset J^\pm({\rm supp}(f)).
\end{align*}
One can introduce the advanced-minus-retarded operator $\Ett = \Ett^- - \Ett^+$ so that
\[
(\Box + m^2)\Ett f = 0 = \Ett(\Box + m^2) f\,, \quad f \in \CoinfM\,,
\]
i.e.\ $\Ett$ maps smooth, compactly supported test-functions to smooth solutions
$\Ett f$ of the Klein-Gordon equation. 

There is the complex conjugation map $\Gamma : C^\infty(M) \to C^\infty(M)$ defined by
$(\Gamma f)(x) = \overline{f(x)}$, $x \in M$. Note that $\Gamma (\Box + m^2) =
(\Box + m^2) \Gamma$ which says that $\Box + m^2$ maps real (resp.\ imaginary)
functions to real (resp.\ imaginary) functions; this property is inherited by
$\Ett^\pm$ and $\Ett$, so that $\Gamma \Ett^\pm = \Ett^\pm \Gamma$,
$\Gamma \Ett = \Ett \Gamma$. Clearly $\Gamma$ extends to a complex conjugation
on $L^2(\Mb)$ with $\langle \Gamma f, \Gamma h \rangle = \langle h, f \rangle$. The map
$\Ett$ can be used to define the {\em causal propagator}, a bi-linear form
$E$ on $\CoinfM \times \CoinfM$ given by
\begin{align}
 E(f,h) & = \langle \Gamma f, \Ett h \rangle \,, \quad f,h \in \CoinfM\,,
 \end{align}
and one can show that 
\begin{align} \label{turn}
E(\Gamma f,  h) & = - E(h,\Gamma f)\,, \quad f,h \in \CoinfM\,.
\end{align}
The continuity properties of $\Ett$ entail that $f \otimes h \mapsto E( f, h)$
extends to a distribution in $\Df'(M \times M)$.

\subsection{Quantized Klein-Gordon field}

Given a globally hyperbolic spacetime $\Mb = (M,g_{ab})$, one can define the
quantized Klein-Gordon field associated to it: This is defined as a family
of symbols $\{ \phi(f) : f \in \CoinfM\}$ which are elements of a $*$-algebra
$\Ff(\Mb)$ fulfilling the following properties:
\begin{itemize}
\item[(i)] $f \mapsto \phi(f)$ is $\mathbb{C}$-linear
\item[(ii)] $\phi(f)^* = \phi(\Gamma f)$
\item[(iii)] $[\phi(f),\phi(h)] = i E(f,h) {\bf 1}$
where $[A,B] = AB -BA$ denotes the commutator and ${\bf 1}$ is the algebraic
unit element in $\Ff(\Mb)$
\item[(iv)] $\phi((\Box + m^2)f) = 0$
\item[(v)] $\Ff(\Mb)$ is algebraically generated by ${\bf 1}$ and the $\phi(f)$, $f \in \CoinfM$.
\end{itemize}
The algebra $\Ff(\Mb)$ will be called the {\em CCR algebra of the quantized Klein-Gordon field} on
$\Mb$.
It can be shown that an algebra $\Ff(\Mb)$ with the listed properties
exists; moreover, keeping $m$ fixed, the algebra $\Ff(\Mb)$ together with the generating elements
$\phi(f)$ is determined uniquely by $\Mb$ up to canonical isomorphism. In consequence, the
assignment $\Mb \to \Ff(\Mb)$ entails a property of local covariance which we discuss a bit
more in detail in Sec.~\ref{sect:loccov}. Note that at this level, $\Ff(\Mb)$ is not represented on any Hilbert space, but we will address this point next.

\subsection{States and representations}

A linear functional $\omega : \Ff(\Mb) \to \mathbb{C}$ is called a {\em state} if $\omega$ is
positive, i.e.\ $\omega(F^*F) \ge 0$ for all $F \in \Ff(\Mb)$, and if $\omega({\bf 1}) = 1$.
Furthermore, one requires that the {\em n-point functions} of $\omega$,
\[
 w^{(n)}_\omega( f_1 \otimes \ldots \otimes f_n) = \omega(\phi(f_1) \cdots \phi(f_n)) \,, \quad
  f_j \in \CoinfM,\ n \in \mathbb{N}\,,
\]
extend by linearity to distributions in $\Df'(M^n)$. 
Any state is uniquely determined by its sequence of $n$-point functions.
However, it should be noted that not every 
such mathematically defined ``state'' corresponds to a physically realistic
configuration of the system. In other words, selection criteria must be imposed and their
consequences and utility must be studied. In fact, it is the main purpose of the present work
to demonstrate that for a specific class of spacetimes, the states on $\Ff(\Mb)$ proposed
by~\cite{AAS} are not compatible with the criterion that physical states ought to be Hadamard states,
and that for this reason (and some others) one shouldn't view the proposal by~\cite{AAS} as a viable
selection criterion for physical states, as will be discussed in Sec.~\ref{sect:slab}.

Anyway, whenever $\omega : \Ff(\Mb) \to \mathbb{C}$ is a ``mathematical'' state, it determines
canonically a Hilbert space representation of $\Ff(\Mb)$, usually called
{\em GNS-representation} of $\omega$. In greater detail, let $\omega: \Ff(\Mb) \to \mathbb{C}$ be
a state. Then there is a collection of objects $(\mathcal{H}_\omega,\pi_{\omega},\mathcal{D}_\omega,\Omega_\omega)$,
where: (1) $\mathcal{H}_\omega$ is a Hilbert space with dense subspace $\mathcal{D}_\omega$,
(2) $\pi_\omega$ is a linear map taking every element $F \in \Ff(\Mb)$ to a closable
operator $\pi_\omega(F)$ defined on $\mathcal{D}_\omega$, (3) $\pi_\omega(F)\mathcal{D}_\omega \subset
\mathcal{D}_\omega$ for all $F \in \Ff(\Mb)$, (4) $\pi_\omega(F F') = \pi_\omega(F)\pi_\omega(F')$
and $\pi_\omega(F)^*|_{\mathcal{D}_\omega} = \pi_\omega(F^*)$ for all $F,F' \in \Ff(\Mb)$, (5) 
$\pi_\omega({\bf 1}) = {\bf 1}$ (the unit operator on $\mathcal{H}_\omega$), (6)
$\Omega_\omega$ is a unit vector in $\mathcal{D}_\omega$ with the property
\[ 
\omega(F) = (\Omega_\omega,\pi_\omega(F) \Omega_\omega ) \,, \quad F \in \Ff(\Mb)\,,
\]
where $(\,.\,,\,.\,)$ denotes the scalar product of $\mathcal{H}_\omega$. The vector $\Omega_\omega$
is called the {\em GNS vector} of $\omega$. Usually, one adopts the minimal choice for
$\mathcal{D}_\omega$: $\mathcal{D}_\omega = \pi_\omega(\Ff(\Mb))\Omega_\omega$.

\subsection{Quasifree states}

A state $\omega$ on $\Ff(\Mb)$ is called {\em quasifree} if it is determined by its
2-point function $w^{(2)}_\omega$ through the requirement that $w^{(n)}_\omega = 0$ for
all odd $n$, and for all even $n = 2 \nu$,
\[
 w^{(2 \nu)}_\omega(f_1 \otimes \ldots f_n) = \sum_p \prod_{k = 1}^\nu w^{(2)}(f_{p(k)},f_{p(k + \nu)})\,,
\]
where the sum runs over all permutations $p$ of $\{1,\ldots,2 \nu\}$ which satisfy
$p(1) < \ldots < p(\nu)$ and $p(k) < p(k + \nu)$. 
\\[6pt]
Because of its prominent appearence in what follows, we shall from now on denote the two-point function
of a quasifree state $\omega$ more simply by
\[
   W_\omega(f,h) = w_\omega^{(2)}(f,h) \,, \quad f,h \in \CoinfM \,.
   \]
${}$ \\
There are a couple of other characterizations
of quasifree states which are useful to studying their properties. First of all, one
observes that  $E(f,h) = E(f',h')$ if $\Ett(f - f') = 0$ or $\Ett(h - h') = 0$.
Therefore, one can define $S(\Mb) = \CoinX{M,\mathbb{R}}/{\rm ker}(\Ett)$, i.e.\ the {\em real-linear} vector space of 
all equivalence classes $[f]$ with respect to the equivalence relation $f \sim f' \Leftrightarrow
\Ett(f - f') = 0$ in $C_0^\infty(M,\mathbb{R})$, the space of real-valued test-functions. Then
it is known that 
\[
 \sigma([f],[h]) = E(f,h)
\]
defines a symplectic form on $S(\Mb)$. Moreover, writing the two-point function $W_\omega$ of
a quasifree state $\omega$ on $\Ff(\Mb)$ in the form 
\begin{align} \label{wmu}
   W_\omega(f,h) & = \mu_\omega([f],[h]) + \frac{i}{2}\sigma([f],[h]) \,, \quad
   f,h \in C_0^\infty(M,\mathbb{R})\,, 
\end{align}
one finds that $\mu_\omega$ is a real-linear scalar product on $S(\Mb)$ which dominates
the symplectic form $\sigma$, meaning that
\begin{align} \label{eq:dominate}
 | \sigma([f],[h])|^2 & \le 4 \mu_\omega([f],[f]) \mu_\omega([h],[h]) \,, \quad [f],[h] \in S(\Mb)\,.
\end{align}
Conversely, if there is a real-linear scalar product  $\mu_\omega$ on $S(\Mb)$ which fulfills
\eqref{eq:dominate}, then it determines via \eqref{wmu} the 2-point function of a quasifree state
$\omega$ --- and hence the quasifree state $\omega$ in full. Thus, there is a one-to-one
correspondence between quasifree states $\omega$ on $\Ff(\Mb)$ and real-linear scalar products
$\mu_\omega$ on $S(\Mb)$ dominating the symplectic form $\sigma$ as in \eqref{eq:dominate}. See
\cite{ManuVerb:1968,KayWald:1991} for further discussion. Note that given a $\mu = \mu_\omega$ with the property 
\eqref{eq:dominate}, one can form the completion $S_\mu = S_\mu(\Mb)$ of $S(\Mb)$ to a real-linear Hilbert space.
One can, moreover, extend $\mu$  and $\sigma$ to $S_\mu$ (denoting the extensions by the same
symbols); relation \eqref{eq:dominate} then remains valid for the extension. Then the extended
$\sigma$ is still an anti-symmetric bilinear form on $S_\mu$. It might fail to be non-degenerate
on $S_\mu$, but owing to \eqref{eq:dominate}, there is a real-linear, bounded 
operator $R_\mu$ with operator norm $\|R_\mu\|\le 1$ on $S_\mu$ such that 
\[
   \sigma(\psi,\psi') = 2\mu(\psi,R_\mu \psi')\,, \quad \psi,\psi' \in S_\mu\,.
\]
The operator $R_\mu$ is called the {\em polarizator} of $\mu$. It fulfills $R_\mu^* = -R_\mu$ where
the adjoint is defined by $\mu(R_\mu^*\psi,\psi') = \mu(\psi,R_\mu \psi')$. It is plain that there
is a one-to-one correspondence between polarizators and real-linear scalar products on $S(\Mb)$
dominating $\sigma$, and hence there is a one-to-one correspondence between polarizators and 
quasifree states. 
The following conditions for a quasifree state $\omega$ on $\Ff(\Mb)$ are known to be equivalent
\cite{ManuVerb:1968,KayWald:1991,ArakiYamagami:1982}.
(a) $\omega$ is pure, i.e.\ it cannot be decomposed as a convex combination of different states.
(b) For $\mu = \mu_\omega$, it holds that $R_\mu^2 = -{\bf 1}$.
(c) The following {\em saturation property} of $\sigma$ holds with respect to $\mu = \mu_\omega$:
\begin{align}
 \mu([f],[f]) & = \sup_{0 \ne [h] \in S(\Mb)}\, \frac{|\sigma([f],[h])|^2}{4 \mu([h],[h])}\,,
 \quad [f] \in S(\Mb)\,.
\end{align}

\section{S-J States}\label{sect:SJ}

In~\cite{AAS}, the authors considered the quantized Klein-Gordon field on a globally hyperbolic
spacetime $\Mb$, and made the following observation. If $\Ett = \Ett^- - \Ett^+$ is
the difference of advanced and retarded fundamental solutions of the Klein-Gordon operator
$\Box + m^2$ on $\Mb$, and if $\Ett f$ is in $L^2(\Mb)$ for each $f \in \CoinfM$,
then 
\[
 A f = i \Ett f\,, \quad f \in \CoinfM\,,
\]
defines a symmetric operator in $L^2(\Mb)$ with domain $\CoinfM$, i.e.\ $\langle f, A h \rangle
= \langle A f, h \rangle$ for all $f,h \in \CoinfM$, as a consequence of \eqref{turn}. In the
case that $\Mb$ is a relatively compact subset of Minkowski spacetime (of dimension $2$ or $4$),~\cite{AAS} showed
that $A$ extends to a bounded operator on $L^2(\Mb)$, and they noted that it is very likely
that $L^2(\Mb)$ boundedness holds also if $\Mb$ is a globally hyperbolic spacetime of these dimensions where $M$ has finite volume with respect to the volume form $d{\rm vol}_{\Mb}$.
A general and rigorous result in this direction can be given as follows in arbitrary spacetime dimension (we sketch the proof in the Appendix):
\begin{Prop}\label{prop:bdedA}
Let $\Mb$ and $\Nb$ be globally hyperbolic spacetimes. Suppose there is an isometric embedding $\psi:\Mb\to\Nb$, preserving orientation and time-orientation, and so that $\psi(M)$ is a causally convex and relatively compact subset of $N$. Then the operator $A$ defined above is a bounded operator on $L^2(\Mb)$.
\end{Prop}

Assuming that $A$ extends to a bounded operator on $L^2(\Mb)$, it possesses a spectral measure
$dP_A$, which can be used to extract the positive part,
\[
 A^+ = \int_{[0,\infty)} \lambda\, dP_A(\lambda) \,.
\] 
Then~\cite{AAS} propose to define a quasifree ``S-J state'' $\omega_{SJ}$ on
$\Ff(\Mb)$, by setting its two-point function to be
\begin{align}
 W_{SJ}(f, h) = \langle \Gamma f, A^+ h \rangle\,, \quad f,h \in \CoinfM\,.
\end{align}
As $A$ is by assumption $L^2(\Mb)$ bounded, so is $A^+$, entailing that $W_{SJ}$ extends
to a distribution in $\Df'(M \times M)$. It was argued in~\cite{AAS} that $W_{SJ}$ really
is the 2-point function of a state on $\Ff(\Mb)$. Let us demonstrate this here in slightly
different form. The first property that needs to be checked is  $W_{SJ}((\Box + m^2)f,h)
= 0 = W_{SJ}(f, (\Box + m^2)h)$ for all $f,h \in \CoinfM$. Let $P^+ = \int_{[0,\infty)} dP_A(\lambda)$
be the spectral projector of $A$ corresponding to non-negative spectral values. Then $A^+ = P^+ A =
A P^+$ and one has for all $f,h \in \CoinfM$,
\[
W_{SJ}(f,(\Box+ m^2)h) = \langle \Gamma f, P^+ i\Ett(\Box + m^2)h\rangle = 0 
\]
since $\Ett(\Box + m^2) = 0$. Similarly one can prove $W_{SJ}((\Box + m^2)f,h) = 0$.

Now let $P^- = {\bf 1} - P^+$, and define $A^- = P^- A = A P^-$ and $|A| = (A^*A)^{1/2}$. Then
\begin{align} \label{Apositive}
 A^+ = \frac{1}{2}(A + |A|)\,, \quad A^- = \frac{1}{2}(A - |A|)\,.
\end{align}
Furthermore, $\Gamma \Ett \Gamma = \Ett$ implies $\Gamma A \Gamma = - A$ and hence
$ \Gamma |A| \Gamma = |A|$, entailing
$\Gamma A^+ \Gamma = - A^-$.
In view of \eqref{Apositive} one obtains for all real-valued
test-functions $f,h \in C_0^\infty(M,\mathbb{R})$,
\begin{align} \label{w2form}
 W_{SJ}(f,h) & = \langle \Gamma f, \frac{1}{2}(|A| + A) h \rangle 
                    = \frac{1}{2} \langle f, |A| h \rangle + \frac{i}{2}\sigma([f],[h]) \,,
\end{align}
where $\sigma([f],[h]) = E(f,h)$. We would like to show that
\begin{align} \label{musj}
   \mu_{SJ}([f],[h]) & = \frac{1}{2} \langle f,|A| h\rangle\,, \quad [f],[h] \in S(\Mb)\,,
\end{align}
defines a real-linear scalar product on $S(\Mb)$ dominating the symplectic form $\sigma$. 
Let $A = U|A|$ be the polar decomposition of (the closure of) $A$. It holds that $U^*U |A| = |A|$ and
therefore, $\Ett(h - h') = 0$ implies $\langle f, |A| (h - h') \rangle = 
\langle U f, A(h - h') \rangle = 0$. Similarly, $\Ett(f - f') = 0$ implies
$\langle (f - f'),|A| h\rangle = 0$. This shows that the right hand side of \eqref{musj} is independent
of the choice of $f$ in $[f]$, resp.\ $h$ in $[h]$, and thus $\mu_{SJ}$ is well-defined on
$S(\Mb)$. We also have for all $f,h \in C_0^\infty(M,\mathbb{R})$,
\[
\overline{ \langle f,|A| h \rangle} = \langle \Gamma f, \Gamma |A| h \rangle = \langle f, |A| h \rangle
\]
since $|A|$ and $\Gamma$ commute and $\Gamma f = f$, $\Gamma h = h$. Thus, $\mu_{SJ}$
takes real values. To see that $\mu_{SJ}$ dominates the symplectic form $\sigma$, it suffices to note
that $\langle f, (|A| - A) f\rangle \ge 0$ for all $f \in \CoinfM$, and to apply the resulting Cauchy-Schwarz
inequality to the bilinear form $\langle f, (|A| - A) h \rangle =
2(\mu_{SJ}([f],[h]) + i \sigma([f],[h])/2)$, $[f],[h] \in S(\Mb)$. This, in turn, shows that 
$\mu_{SJ}$ is non-degenerate (since so is $\sigma$), finally proving that $\mu_{SJ}$ is a real
scalar product on $S(\Mb)$ dominating $\sigma$.

We note that these arguments remain valid even if $A$ is not bounded or essentially self-adjoint. Once
$\Ett f$ is in $L^2(\Mb)$ for all $f \in \CoinfM$,
$A$ is symmetric on $\CoinfM$, hence it is closable and its closure has a unique polar decomposition
$A = U|A|$; the 2-point function can still be defined via \eqref{w2form}, and the arguments 
showing that the  $\mu_{SJ}$ of \eqref{musj} yields a real scalar product on $S(\Mb)$ dominating
$\sigma$ continue to hold without change. The point which does not generalize immediately is the
property of $W_{SJ}$ to be a distribution in $\Df'(M \times M)$. This might also be shown but
we won't do this here as the $A$ we will study in the next section will turn out to be
$L^2(\Mb)$ bounded. For this situation --- $A$ bounded and hence, in particular, self-adjoint ---
we can state the following result.
\begin{Prop}\label{prop:pureqf}
$W_{SJ}$ is the two-point function of a pure quasifree state on $\Ff(\Mb)$.
\end{Prop}
{\em Proof}. Note that $A = U |A| = |A|U^*$ in the polar decomposition with some partial isometry $U$. Since
$A$ is self-adjoint, $U$ and $|A|$ commute. Since $|A|$ commutes with $\Gamma$ while $A$ and 
$\Gamma$ anti-commute, $U$ and $\Gamma$ anti-commute. 
In other words, the operator $iU$ on $L^2(\Mb)$ commutes with $\Gamma$ and hence maps
real-valued functions to real-valued functions.
Let $f \in C_0^\infty(M,\mathbb{R})$. 
Then there is a sequence $h_n$ in $C_0^\infty(M,\mathbb{R})$ which
converges to $iUf$. We have
$$ \frac{|\sigma([f],[h_n])|^2}{4 \mu_{SJ}([h_n],[h_n])} =
 \frac{|\langle f,A h_n\rangle|^2}{4 \langle h_n,|A| h_n\rangle/2}\,;$$
the numerator converges to $|\langle f, i AU f \rangle|^2 = 
|\langle U^*U|A|f,f\rangle|^2 = |\langle f,|A| f \rangle|^2$ since $U^*U = 1$ on ${\rm range}(|A|)$.
The denominator converges to $ 2 \langle iU f,i|A|Uf \rangle =
2 \langle  f, U^*U |A| f \rangle = 2\langle f,|A| f \rangle$ .
This means we have for any $[f] \in S(\Mb)$,
$$\lim_{n \to \infty} \frac{|\sigma([f],[h_n])|^2}{4 \mu_{SJ}([h_n],[h_n])} = \frac{1}{2}\langle
 f,|A| f\rangle = \mu_{SJ}([f],[f])\,,$$
which establishes the required saturation property of $\sigma$ with respect to $\mu_{SJ}$.
${}$ \hfill $\Box$

\section{S-J States on an Ultrastatic Slab Spacetimes}\label{sect:slab}

\subsection{Spectral representation of $W_{SJ}$}

Let $\Sigma$ be a compact $d$-manifold with smooth Riemannian metric $h$. 
We will consider the S-J prescription on ultrastatic slab spacetime $\Mb$ with
manifold $M=(-\tau,\tau)\times\Sigma$ and metric $ds^2 = dt^2 - h_{ij} dx^i dx^j$
for $\tau>0$. On such a spacetime, the Klein--Gordon equation becomes
\[
\left(\frac{\partial^2}{\partial t^2} +K\right)\phi = 0\,,
\]
where $K = -\triangle + m^2$, $\triangle$ being the Laplacian on $(\Sigma,h)$. We
keep $m>0$ to avoid problems with zero-modes.\footnote{Much of our analysis applies if
$K$ is any strictly positive elliptic partial differential operator with smooth real coefficients.}
Then $K$ is essentially self-adjoint on $\CoinX{\Sigma}\subset L^2(\Sigma)$, defined using the volume measure induced by $h$. The unique self-adjoint extension will be
denoted $K$ and will be assumed to be strictly positive. There is a complete orthonormal basis for $L^2(\Sigma)$ of $K$-eigenfunctions $\psi_j$ ($j\in J$) such that $K\psi_j = \omega_j^2\psi_j$ with each $\omega_j>0$. (These facts
are standard, but for a reference, see~\cite[Ch.~8]{Taylor_volII}.)
The index set $J$ is countable, of course, and we make the assumption that $\overline{\psi_j}$ is also one of the 
basis elements for each $j$, which simplifies certain formulae. (One could allow $\Sigma$
to have boundary, at the expense of possibly needing to select a particular self-adjoint extension 
of $K$.) The spacetime Hilbert space is $L^2(\Mb) = L^2(-\tau,\tau)\otimes L^2(\Sigma)$. 

The advanced-minus-retarded fundamental solution on $M$ may be expanded using the basis $\psi_j$;
it is just the restriction to $M$ of the corresponding bidistribution on $\RR\times\Sigma$, namely
\[
E(t,\ux;t',\ux') = \sum_{j\in J} \frac{\sin\omega_j (t'-t)}{\omega_j} \psi_j(\ux)\overline{\psi_j(\ux')}.
\]
Proceeding according to the S-J prescription, we define the operator $A  =i\Ett$; this is bounded by
Prop.~\ref{prop:bdedA} (by embedding our ultrastatic slab in $\RR\times\Sigma$) or by direct
computation as below. To analyse $A$ it is convenient to proceed as follows. 
For each $j\in J$ there is an isometry $U_j: L^2(-\tau,\tau)\to L^2(M)$ given by 
$U_j f = f\otimes\psi_j$, such that $K U_j^* =\omega_j^2 U_j^*$ for all $j$. 
Then the operator $A = i\Ett$ may be written as a sum
\begin{equation}\label{eq:Asum}
A = \sum_{j\in J} U_j A_j U_j^*\,,
\end{equation}
where $A_j$ is the operator on $L^2(-\tau,\tau)$ 
\[
A_j f = \frac{i}{\omega_j} \left(\ip{S_j}{f}C_j - \ip{C_j}{f}S_j\right) 
\]
and $C_j(t)= \cos\omega_j t$, $S_j(t)= \sin\omega_j t$ as elements of $L^2(-\tau,\tau)$. 
To be precise, we need to specify the sense in which the sum converges: we use the 
weak operator topology, i.e.,~\eqref{eq:Asum} means that
\[
\ip{f}{Ah} = \sum_{j\in J} \ip{f}{U_j A_j U_j^* h}
\]
for each $f,h\in L^2(\Mb)$ (with the sum converging absolutely in $\CC$).\footnote{In fact, 
with the particular operator $A=i\Ett$,~\eqref{eq:Asum} even converges in the
strong operator topology.} As a number of series of the form~\eqref{eq:Asum} will appear below, 
we note that any such sequence will converge in the weak operator topology (to a bounded limit)
provided that $\sup_{j\in J}\|A_j\|<\infty$,
whereupon $\|A\|$ is the value of this supremum. Our main
interest will be where the $A_j$ are finite-rank self-adjoint operators; in this case,
the limit is self-adjoint, and there is an orthonormal basis for $L^2(\Mb)$ of eigenvectors formed from those of the $A_j$ in an obvious way. 

In the case $A=i\Ett$, each operator $A_j$ has rank $2$, so its eigenvectors are easily found.  
Computing some inner products, 
\[
\ip{C_j}{S_j} = 0,\qquad \|C_j\|^2=  \tau (1+\sinc 2\omega_j\tau),\qquad
\|S_j\|^2=  \tau (1-\sinc 2\omega_j\tau),
\]
where $\sinc x = \sin x/x$. As $C_j$ and $S_j$ are orthogonal, we find
\[
A_jS_j = \frac{i}{\omega_j} \|S_j\|^2 C_j, \qquad A_jC_j = -\frac{i}{\omega_j} \|C_j\|^2 S_j
\]
and therefore deduce that the eigenvalues of $A_j$ are $\pm \|S_j\| \|C_j\|/\omega_j$ with 
eigenvectors
\[
\varphi_j^\pm = C_j \mp i \frac{\|C_j\|}{\|S_j\|} S_j 
\]
or, 
\[
\varphi_j^\pm(t) =  e^{-i\omega_j t} + i\left(1\mp \frac{\|C_j\|}{\|S_j\|}\right) 
\sin \omega_j t .
\]
We therefore see directly that $\|A_j\|\le 2\tau/\omega_j$ is uniformly bounded in $j$,
so $A$ is bounded and an orthonormal basis of  eigenvectors may be formed from those of the $A_j$. 

Owing to the orthogonality of $C_j$ and $S_j$, $\|\varphi_j^\pm\|^2 = 2\|C_j\|^2$. 
The positive part of $A_j$ is now seen to be
\[
A_j^+ =  \frac{\|S_j\| \|C_j\|}{\omega_j}\, \frac{\varphi_j^+\bra{\varphi_j^+}}{2\|C_j\|^2}
\]
or, as an integral kernel,
\[
A_j^+(t,t') = \frac{\|S_j\|}{2\omega_j \|C_j\|} \left(
e^{-i\omega_j t} + i \delta_j \sin \omega_j t \right)
 \left(
e^{i\omega_j t'} - i \delta_j \sin \omega_j t'\right),
\]
where $\delta_j= 1 -\|C_j\|/\|S_j\|$. For future reference, we note that
\[
\delta_j = 1-\frac{\|C_j\|}{\|S_j\|} = 1 - \sqrt{1 + \frac{2\sinc 2\omega_j \tau}{1-\sinc 2\omega_j \tau}}
 = -\sinc 2\omega_j \tau + O((\omega_j\tau)^{-2})
\]
and
\[
\varphi_j^+ = \left(1-\frac{\delta_j}{2}\right) e^{-i\omega_j t} + \frac{\delta_j}{2}e^{i\omega_j t}.
\]

The positive part of the operator $A$ is simply 
\[
A^+ = \sum_{j\in J} U_j A_j^+ U_j^*,
\]
with the sum again converging in  the weak operator topology. This allows us to identify the
two-point function of the S-J state as 
\[
W_{SJ}(t,\ux;t',\ux') = \sum_{j\in J} 
\frac{\|S_j\|}{2\omega_j \|C_j\|}
\left(
e^{-i\omega_j t} + i \delta_j \sin \omega_j t \right)
 \left(
e^{i\omega_j t'} - i \delta_j \sin \omega_j t'\right) \psi_j(\ux)\overline{\psi_j(\ux')},
\]
where the convergence of the sum is understood in a distributional sense, i.e.,
the smeared two-point function $W_{SJ}(f,h)$ is obtained by integrating
each term in the sum against $f(t,\ux)h(t',\ux')$ and then performing the sum on $j$. 
As already shown, this gives a distribution that is the two-point function
of a pure quasifree state. 

\subsection{Failure of Hadamard property for $W_{SJ}$}

In this section we prove that $W_{SJ}$ on $M = (-\tau,\tau)\times\Sigma$ fails to be Hadamard except (at most) a set of $\tau$ with measure zero. We also give two standard examples 
in which there is no value of $\tau$ for which the S-J state is Hadamard.

It will be useful to have a Hadamard two-point function for comparison. 
This is provided by the restriction to $M$ of the two-point function of the usual ground state 
on $\RR\times\Sigma$, 
\[
W_H (t,\ux;t',\ux') = \sum_{j\in J} \frac{e^{-i\omega_j (t-t')}}{2\omega_j}\psi_j(\ux)\overline{\psi_j(\ux')},
\]
with the sum understood in the distributional sense, and 
which is known to be Hadamard~\cite{FullingNarcowichWald}. Restricted to the ultrastatic slab, 
$W_H$ can also be obtained from a self-adjoint bounded operator on $L^2(\Mb)$, namely 
\[
A^{(H)} = \sum_{j\in J} U_j A_j^{(H)} U_j^*,
\]
where $A_j^{(H)}$ is the rank-$1$ operator
\[
(A_j^{(H)} f)(t)= \frac{e^{-i\omega_j t}}{2\omega_j}\int_{-\tau}^\tau e^{i\omega_j t'} f(t') \,dt'
\]
for $f\in L^2(-\tau,\tau)$. Unless $\sinc 2\omega_j \tau = 0$ for all $j$, 
we see that the S-J state is not simply the restriction of the standard ground state (thus exemplifying the remarks at the end of Sect.~\ref{sect:SJ}). 

Let us suppose $\tau$ is such that $\omega_{SJ}$ is Hadamard. Then the normal ordered two-point function ${:}W_{SJ}{:} = W_{SJ}-W_H$ is smooth on $M\times M$; in particular it should be
$C^2$ on $M\times M$, and, on further restriction to any subset $M'=(-\tau',\tau')\times\Sigma$ 
with $0<\tau'<\tau$, the derivatives may be assumed to be bounded on $M'\times M'$. 
In particular,
\[
F(t,\ux;t',\ux')  = \frac{\partial^2}{\partial t\partial t'} {:}W_{SJ}{:}(t,\ux;t',\ux') 
\]
should be square-integrable on $M'\times M'$, i.e., $F\in L^2(\Mb'\times \Mb')$, which
means that $F$ is the integral kernel of a Hilbert--Schmidt operator 
$T$ on $L^2(\Mb')$. Now the subtraction and differentiations used to compute $F$ may be performed
mode-by-mode; for example, ${:}W_{SJ}{:}$ has distributional kernel
\begin{equation}\label{eq:WS_normord}
{:}W_{SJ}{:}(t,\ux;t',\ux') = \sum_{j\in J} \left\{\frac{\delta_j^2\cos\omega_j (t-t')}{4\omega_j(1-\delta_j)}
+\frac{\delta_j(2-\delta_j)}{4\omega_j(1-\delta_j)}\cos\omega_j(t+t')\right\}
\psi_j(\ux)\overline{\psi_j(\ux')}.
\end{equation}
The mode-by-mode subtraction is justified because ${:}W_{SJ}{:}$ is actually
the operator kernel of $A^+-A^{(H)}$, which may be written as the
weakly converging sum $\sum_{j\in J} U_j (A_j^+-A_j^{(H)}) U_j^*$. By
similar reasoning, the operator $T$ on $L^2(\Mb')$ may be written in the form
\[
T =  \sum_{j\in J} U'_j T_j (U'_j)^*,
\]
where $T_j$ is the operator on $L^2(-\tau',\tau')$ 
\[
T_j f = \frac{\omega_j\delta_j}{2} \left(\frac{1}{1-\delta_j}\ip{S_j}{f} S_j - \ip{C_j}{f} C_j \right)
\]
with $C_j$ and $S_j$ now restricted to $L^2(-\tau',\tau')$,
and the isometry $U_j':L^2(-\tau',\tau')\to L^2(M')$ defined by $U'_jf = f\otimes\psi_j$. 
Each $T_j$ is a rank-$2$ self-adjoint operator with eigenvalues $-\omega_j\delta_j \|C_j\|^2/2$ and
$\omega_j\delta_j\|S_j\|^2/(2(1-\delta_j))$, where the norms $\|C_j\|$ and $\|S_j\|$
are taken in $L^2(-\tau',\tau')$, but other quantities are defined exactly as before. 
Accordingly, $T$ has eigenvalues
\[
\left\{ -\omega_j\delta_j \|C_j\|^2/2: j\in J\right\} \cup 
\left\{\omega_j\delta_j\|S_j\|^2/(2(1-\delta_j)) :  j\in J\right\} \cup\{0\},
\]
and, as the eigenvalues of a Hilbert-Schmidt operator are square-summable, we have
\begin{equation}\label{eq:nec}
\sum_{j\in J} \omega_j^2\delta^2_j \|C_j\|^4 <\infty, \qquad\text{and}\qquad
\sum_{j\in J} \frac{\omega_j^2 \delta^2_j \|S_j\|^4}{(1-\delta_j)^2} <\infty.
\end{equation}
Hence $\omega_j\delta_j\to 0$ in some (and hence any) ordering of $J$ by the
natural numbers. In what follows, we assume that the $\omega_j$ have 
been ordered so that $\omega_j\le \omega_{j+1}$.\footnote{If $K$ has degenerate eigenspaces then
there may be finitely many $j$ with the same value of $\omega_j$.}
Given that $\omega_j \delta_j   \sim (2\tau)^{-1}\sin 2\omega_j\tau$, we have proved
the following. 

\begin{Prop} There exists $0<\tau'<\tau$ for which ${:}W_{SJ}{:}\in C^2(M'\times M')$
only if $\sin 2\omega_j\tau\to 0$ as $j\to\infty$. 
\end{Prop}
We emphasise that this is only a necessary condition. Let $V$ be the set of $\tau$ for which it holds. 
Then $V$ may easily be seen to be a Borel subset of $\RR$; but, as we now show, it has zero Lebesgue measure.
For suppose otherwise, and assume without loss of generality that $V$ has nonzero, finite measure.\footnote{If $V$ has infinite measure, consider intersections with intervals of the form $(0,T)$.} Then its characteristic
function $\chi$ is an $L^1$ function such that 
$\chi(\tau)\sin^2 2\omega_j\tau \to 0$ as $j\to\infty$ for all $\tau$, so the dominated convergence theorem gives
\[
\int_{0}^\infty \chi(\tau) \sin^2 2\omega_j\tau \,d\tau \to 0\qquad \text{as $j\to\infty$}.
\]
However, by elementary trigonometric identies, the integrand is $\frac{1}{2}
\chi(\tau)(1-\cos 4\omega_j\tau)$, and the Riemann--Lebesgue lemma entails that
\[
\int_0^\infty \chi_T(\tau)\cos 4\omega_j\tau\,d\tau \to 0\qquad \text{as $j\to\infty$}.
\]
We conclude that $\int_0^\infty \chi(\tau)\,d\tau=0$, i.e.,  the Lebesgue measure of $V$ 
vanishes, and we obtain a contradiction. In summary, we have proved (somewhat more than) 
the following.
\begin{Thm} 
The set of $\tau$ for which the S-J state is Hadamard on $\Mb$ is contained
in a set of measure zero.
\end{Thm}
To conclude this subsection, we give two examples in which it is easy to see directly that $V$ is empty, which means there is {\em no} choice of $\tau$ for which the S-J state is Hadamard. 

\paragraph{Toroidal spatial section} Suppose $K=-\triangle+m^2$, where $\triangle$ is the Laplacian on a flat $3$-torus with common periodicity length $L$, i.e., $\Sigma = \RR^3/(L\ZZ)^3$, and $m>0$ to avoid zero modes. Then 
\[
\omega_{\kb} = \sqrt{(2\pi\|\kb\|/L)^2  +m^2},
\]
where $\kb\in\ZZ^3$. In particular, the subsequences $\omega_{(r,0,0)}$ and $\omega_{(r,r,0)}$ 
obey $\omega_{(r,0,0)}\sim 2\pi r/L$ and $\omega_{(r,r,0)}\sim 2\pi r\sqrt{2}/L$ as $r\to\infty$. 
The S-J state is Hadamard only if both $\sin 4\pi r\tau/L$ and $\sin 4\sqrt{2}\pi r\tau/L$ converge
to zero as $r\to\infty$. But the first occurs only for $4\tau/L\in \ZZ$, while the second requires
$4\sqrt{2}\tau/L\in \ZZ$, so there is no value of $\tau>0$ for which the S-J state is Hadamard. 

\paragraph{Spherical spatial section} Suppose $K=-\triangle+m^2$, where $\triangle$ is the Laplacian on the round $3$-sphere of radius $R$ and $m>0$. Then 
\begin{align*} 
\omega_j = \sqrt{j(j +2)/R^2 + m^2}, \qquad j=0,1,2,\ldots,
\end{align*}
each level appearing with multiplicity $(j+1)^2$. Evidently $\sin 2\omega_j\tau\to 0$ only if $2\tau/(\pi R)\in\NN$, so the S-J state can be Hadamard only if $2\tau =k\pi R$, for some $k\in  \NN$. 
In such cases, we have
\begin{equation}\label{eq:sin2asy}
\sin^2 2\omega_j\tau\sim \left(\frac{((mR)^2-1)\pi k}{2j}\right)^2
\end{equation}
and the sums in~\eqref{eq:nec} are then easily seen to diverge owing to the multiplicities
of the eigenvalues of $-\triangle$. Accordingly, there is again no value of $\tau>0$ 
for which the S-J state is Hadamard. 

\subsection{Digression: the limit $\tau\to\infty$}

As a slight departure from our main theme, we point out that one can make
rigorous sense of the limit in which the time span $\tau$ is taken to infinity,
as a distributional limit of the two-point function. Thus, for each fixed
$f, h\in\CoinX{\RR\times \Sigma}$, we attempt to define
\[
W^{(\infty)}_{SJ}(f,h) = \lim_{\tau\to\infty} W^{(\tau)}_{SJ}(f,h),
\]
where $W^{(\tau)}_{SJ}$ is the two-point function defined on the slab
with parameter $\tau$ (which, for all sufficiently large $\tau$, contains
the supports of both test functions). The above limit exists if and only if
the analogous limit of the normal ordered two-point functions does, because $W_H$ is independent of $\tau$, and a simple calculation based on Eq.~\eqref{eq:WS_normord} shows
that
\begin{equation}\label{eq:SJ_infinity}
{:}W^{(\infty)}_{SJ}{:}(f,h) =\lim_{\tau\to\infty} 
\sum_{j\in J} \frac{\delta_j}{2\omega_j}
\left(\frac{(\Cc f_j)(\omega_j)(\Cc h_j)(\omega_j)}{1-\delta_j}
 + (\Sc f_j)(\omega_j)(\Sc h_j)(\omega_j)\right),
\end{equation}
where $\Cc$ and $\Sc$ are two-sided cosine and sine transforms
\[
(\Cc f)(\omega)=\int_{-\infty}^\infty f(t)\cos\omega t\,dt, \qquad
(\Sc f)(\omega)=\int_{-\infty}^\infty f(t)\sin\omega t\,dt 
\]
and we write $f_j=U_j^*f$. Now $\delta_j \to 0$ as $\tau\to\infty$ 
for each fixed $j\in J$, and a dominated convergence
argument (see below) shows that this pointwise limit entails
that the series vanishes in this limit. Accordingly, we see that
the limiting two-point function exists, and is precisely the
Hadamard two-point function $W_H$. 

This argument makes precise a similar discussion in~\cite{AAS} (where 
the more general static case is treated): the S-J prescription reproduces, by
the limiting argument, the standard ground state. However, we emphasise, on the one hand, on that the
S-J prescription is not itself well-defined on the whole ultrastatic spacetime,
because $\Ett$ does not map test functions into $L^2$, so its existence even as
an unbounded operator seems unlikely; and, on the other hand, that
the states obtained by the S-J prescription at finite $\tau$ are not Hadamard.

To conclude this subsection, we sketch the dominated convergence argument
mentioned above. General results on the distribution of eigenvalues
for the Laplacian~\cite[Ch.~8, \S 3]{Taylor_volII} show that 
$\sum_{j\in J} 1/(\omega_j(\omega_j^2+m^2)^{2N})$ 
converges for all sufficiently large $N\in\NN$, and we choose any such value. 
If $f\in\CoinX{\RR\times\Sigma}$ then 
$(m^2+\omega^2)^N(\Cc f_j)(\omega)$ is the $L^2$-inner product
of $\psi_j(\ux)\cos\omega t /(1+(m\tau)^2)$ and
$(1+(m\tau)^2)(m^2-\partial^2/\partial t^2)^N f(t,\ux)$, both
of which are square-integrable. By the
Cauchy--Schwarz inequality, one finds that $|(\Cc f_j)(\omega_j)|\le
\text{const}/(\omega_j^2+m^2)^{N}$; a similar result holds
for the sine transform, and thus the summands in Eq.~\eqref{eq:SJ_infinity}
are dominated by those of an absolutely convergent series as
required to apply the dominated convergence theorem.

\subsection{Failure of SJ-Hadamard unitary equivalence for the spherical spatial sections}

For the case of the round $3$-sphere as spatial section, we will now show that, except for $\tau$ in a set of measure zero, the states $\omega_{SJ}$
and $\omega_H$ induce unitarily inequivalent GNS representations $\pi_{SJ}$ and $\pi_{H}$;
that is, there is no unitary $V : \mathcal{H}_{SJ} \to \mathcal{H}_H$ such 
that $V \pi_{SJ}(F) = \pi_H(F) V$ for all $F \in \Ff(\Mb)$. As both states are pure, 
and hence have irreducible representations, this immediately tells us that they are 
{\em disjoint}: no density matrix state induced by $\pi_{SJ}$ arises as a density matrix state induced by $\pi_{H}$, and vice versa. This is the extreme opposite of the
condition of {\em quasiequivalence}, under which every density matrix state induced by 
$\pi_{SJ}$ arises as density matrix states induced by $\pi_{H}$, and vice versa. 
For a detailed discussion of the need of (local) quasiequivalence among GNS representations of states which
are regarded as physical states in quantum field theory in curved spacetime, and the 
consequences of its failure, see \cite{Wald_qft}. Disjoint states induce
incompatible particle number operators (cf.\ Sec.\ 5.2.3 in \cite{BratRob2}) and therefore,
incompatible concepts of particles.

\begin{Prop} \label{prop:disjoint-H-SJ}
For the GNS representations $\pi_{SJ}$ and $\pi_H$ to be unitarily equivalent
it is necessary that $2\tau/(\pi R)\in\NN$. (We make no claim that this condition is sufficient.)
\end{Prop}
{\em Proof}.  We denote by $\mu_{SJ}$ and $\mu_H$ the real scalar products on $S(\Mb)$ corresponding to the
2-point functions $W_{SJ}$ and $W_H$. Then we denote by $S_{SJ}$ and $S_H$ the respective completions;
$\sigma_{SJ}$ and $\sigma_{H}$ will be the continuous extension of the symplectic form $\sigma$.
A necessary condition for unitary equivalence of $\pi_{SJ}$ and $\pi_{H}$ is that $\mu_{SJ}$ and
$\mu_H$ induce the same topology on $S(\Mb)$ \cite{ArakiYamagami:1982, Verch:1994}, implying that $S_{SJ}$
coincides with $S_H$ as topological vector space, and $\sigma_{SJ}
= \sigma_H$. We proceed by assuming that this condition is fulfilled --- otherwise
the proof of the Proposition is already done.
Then it can be seen from \cite{ArakiYamagami:1982,Verch:1994} that a further {necessary} condition is that
$R_H R_{SJ}^{-1} - {\bf 1}$ must be Hilbert-Schmidt on $S_H$,
where $R_H$ and $R_{SJ}$ are the polarizators of $\mu_H$ and $\mu_{SJ}$.
For any $[f],[h] \in S(\Mb)$, it follows that
\begin{align*}
\mu_H([f], ( R_H R_{SJ}^{-1} - {\bf 1}) [h]) & = \mu_{SJ}([f],[h]) - \mu_{H}([f],[h]) = W_{SJ}(f,h) - W_H(f,h) \\
              & = {:}W_{SJ}{:}(f,h) \,.
\end{align*}
Our aim now is to establish existence of an orthonormal system $[e_j]$, $j \in J$, of
$S_H$ such that
\[
{:}W_{SJ}{:} (e_j,e_j) = \mu_H([e_j], ( R_H R_{SJ}^{-1} - {\bf 1}) [e_j])
\]
is not square summable except for $\tau$ in a set of measure zero. This implies that $R_H R_{SJ}^{-1} - {\bf 1}$,
is not
Hilbert-Schmidt in $S_H$ (except for $\tau$ in a set of measure zero).\footnote{
Recall that an operator $T$ is Hilbert--Schmidt if and only if its matrix elements 
$\ip{\zeta_i}{T\zeta_{i'}}$ in some orthonormal basis $\zeta_i$ are square summable.}


In order to construct such an orthonormal system,
 it is useful to label the multiplicities of the eigenvalues
explicitly, for the case at hand where $\Sigma$ is the 3-sphere of radius $R >0$.
We write 
$$ j =({\rm j},\gamma) $$
where ${\rm j} \in \mathbb{N}_0$ and $\gamma$ takes values in $N_{\rm j} =
\{1,2,\ldots,({\rm j} + 1)^2\}$. The eigenvalues of $K$ are
\begin{align} \label{sphere-eigenv}
\omega_j  = \omega_{\rm j} = \sqrt{{\rm j}({\rm j} +2)/R^2 + m^2} \,, 
\end{align}
as before.

Since $ \Gamma K \Gamma = K$, the eigenspaces  of $K$ 
for each eigenvalue are $\Gamma$-invariant,
and we can choose an orthonormal basis $\psi_j = \psi_{({\rm j},\gamma)}$ of
real-valued eigenvectors in $C^\infty(\Sigma,\mathbb{R}) \subset L^2(\Sigma)$ (which also have
compact support, since $\Sigma$ is compact). 

Moreover, for each ${\rm j} \in \mathbb{N}_0$, we can choose some real-valued
$\eta_{\rm j} \in C_0^\infty(-\tau,\tau)$ which additionally is symmetric, i.e.\
$\eta_{\rm j}(t) = \eta_{\rm j}(-t)$, and has the property that
\begin{align} \label{norm_f_j}
 \frac{1}{2\omega_{\rm j}} & \left( \int_{-\tau}^\tau \eta_{\rm j}(t) \cos(\omega_{\rm j} t)\,dt \right )^2 = 1\,.
\end{align}
Note that $\int \eta_{\rm}(t) \cos(\omega_{\rm j} t) \, dt = \sqrt{2 \pi} \widehat{\eta_{\rm j}}(\omega_{\rm j})$
where the hat denotes Fourier transform, on account of symmetry of $\eta_{\rm j}$.

With these assumptions, we define
\begin{align*} 
 e_j = e_{({\rm j},\gamma)} = \eta_{\rm j} \otimes \psi_{({\rm j},\gamma)} \,, \quad
  {\rm j} \in \mathbb{N}_0, \ \gamma \in N_{{\rm j}} \,.
\end{align*}
The $e_j$ are in $C_0^\infty((-\tau,\tau) \times \Sigma,\mathbb{R})$. Furthermore,
inserting them in $W_H$ yields
\[
W_H(e_j, e_{j'}) = \frac{1}{2\omega_j} \int \eta_{\rm j}(t) \cos(\omega_j t) dt
   \int \eta_{{\rm j}'}(t') \cos(\omega_{j'} t') dt' \,\delta_{j j'} = \delta_{j j'} \,,
\]
observing the normalization \eqref{norm_f_j}; the $\delta_{j j'}$ appearing
here is the Kronecker $\delta$. Thus, the $e_j$ form an orthonormal
system with respect to $W_H$, and in view of $W_H(f,h) = 
\mu_H([f],[h]) + i \sigma([f],[h])/2$, it follows that  the $[e_j]$ form an orthonormal
system in $S(\Mb)$ with respect to $\mu_H$. 

Inserting the $e_j$ in ${:} W_{SJ} {:}$ gives, after straightforward calculation,
\[
 {:} W_{SJ} {:} (e_j, e_j) = \frac{\delta_j}{1 - \delta_j}\,.
\]
Now $1 - \delta_{j} $ converges to 1 as ${\rm j} \to \infty$
while, as seen previously, 
\begin{align} \label{delta-asympt}
\delta_{ j} \sim - \sin(2 \omega_{\rm j}\tau)/(2 \omega_{\rm j}\tau)
+ O((\omega_{\rm j} \tau)^{-2}) \,. 
\end{align}
Thus, in order to have 
square summability of ${:} W_{SJ} {:} (e_j, e_j)$ we need convergence of the sum
\[
 \sum_{{\rm j};\gamma}
  \left( \frac{\sin(2\omega_j\,\tau)}
   {2\omega_j \tau} \right)^2\,,
\]
with $\omega_j = \omega_{\rm j}$ as in \eqref{sphere-eigenv}, and $\gamma$ summing over the
multiplicities.
For this to hold it is necessary that $\sin(2\omega_{\rm j}\tau)$ converges
to $0$ as ${\rm j} \to \infty$, but as we have already concluded above, this 
possible only if $2\tau/(\pi R)\in\NN$. 
Accordingly, $R_H R_{SJ}^{-1} - {\bf 1}$ cannot be Hilbert--Schmidt 
in $S_H$ except for the stated (measure zero) set of $\tau$, which completes the proof.  
${}$ \hfill $\Box$
\subsection{Failure of unitary equivalence of S-J states for different values of
$\tau$ in the spherical case}
We shall now demonstrate that S-J states for different time-intervals are generally disjoint.
So let $\tau > \tau' > 0$ and define the ultrastatic slab spacetimes 
$\Mb_\tau = (-\tau,\tau) \times \Sigma$ and $\Mb_{\tau'} = (-\tau',\tau') \times \Sigma$
with ultrastatic metric, where $\Sigma$ is, as in the previous section, the round $3$-sphere
with radius $R > 0$, and let $\Ff(\Mb_{\tau})$ and $\Ff(\Mb_{\tau})$ be
the corresponding algebras for the quantized theory, with corresponding
S-J states $\omega_{SJ\tau}$ and $\omega_{SJ\tau'}$ and associated
two-point functions $W_{SJ,\tau}$, $W_{SJ,\tau'}$.

Now, the local covariance of the Klein--Gordon theory (see Sec.\ \ref{sect:loccov}) 
entails that the canonical isometric embedding $\psi_{\tau,\tau'} : \Mb_{\tau'} \to \Mb_\tau$ given by the identity map $(t,\ux) \mapsto (t,\ux)$ induces an injective $*$-algebra homomorphism $\alpha_{\tau,\tau'} : \Ff(\Mb_{\tau'}) \to \Ff(\Mb_\tau)$, with the action
\begin{equation}\label{eq:alpha_action}
\alpha_{\tau,\tau'} \phi_{\tau'}(f) =\phi_\tau(\psi_{\tau,\tau' *} f)
\end{equation}
and mapping the unit element of $\Ff(\Mb_{\tau'})$ to that of $\Ff(\Mb_{\tau})$,
where the subscript on the Klein--Gordon field indicates which spacetime is meant.
Furthermore, as $\psi_{\tau,\tau'}(\Mb_{\tau'})$ obviously contains a Cauchy-surface 
for $\Mb_\tau$, the map $\alpha_{\tau,\tau'}$ actually is an isomorphism (this is the `time-slice property'~\cite{BrFrVe03}). Thus the S-J state $\omega_{SJ\tau}$
induces a state $\omega_{SJ\tau}\circ  \alpha_{\tau,\tau'}$ on $\Ff(\Mb_{\tau'})$,
which is pure because  $\alpha_{\tau,\tau'}$ is an isomorphism. Now Eq.~\eqref{eq:alpha_action} entails that the $n$-point functions of $\omega_{SJ\tau}\circ  \alpha_{\tau,\tau'}$ are just the distributional pull-backs by
$\psi_{\tau,\tau'}$ of those of $\omega_{SJ\tau}$, which amounts simply
to their restriction to $\Mb_{\tau'}$. In particular, 
$\omega_{SJ\tau}\circ  \alpha_{\tau,\tau'}$ is quasifree and corresponds
to a scalar product $\mu_{SJ\tau}$ on $S(\Mb_{\tau'})$ given by
$\mu_{SJ\tau}([f],[h]) = W_{SJ,\tau}(\psi_{\tau,\tau'*} f,\psi_{\tau,\tau'*} h)$,
which may be written simply as $W_{SJ,\tau}( f, h)$ by a slight abuse of notation.
This can be compared with the scalar product $\mu_{SJ,\tau'}$ obtained
directly from the S-J prescription on $\Mb_{\tau'}$, to gain information
on the GNS representations $\pi_{SJ\tau}$ and $\pi_{SJ\tau'}$ of $\Ff(\Mb_{\tau'})$ 
induced by $\omega_{SJ\tau}\circ  \alpha_{\tau,\tau'}$ and $\omega_{SJ\tau'}$,
with the following result:

%
%

\begin{Prop} \label{disaster}
The set of pairs of values $\tau > \tau' > 0$ such that $\pi_{SJ\tau}$ and $\pi_{SJ\tau'}$ are unitarily
equivalent is contained in a set of measure zero (in $\mathbb{R}^2$).
\end{Prop}
{\it Proof.} The proof follows the same pattern as that of Proposition \ref{prop:disjoint-H-SJ}.
We will assume that $\mu_{SJ\tau}$ and $\mu_{SJ\tau'}$ induce the same
topology on $S(\Mb_{\tau'})$, and we will show that there is an orthonormal system $[e_j]$, $j \in J$,
in $S(\Mb_{\tau'})$
with respect to $\mu_{SJ\tau}$ such that the sequence of differences 
$\mu_{SJ\tau}([e_j],[e_j]) - \mu_{SJ\tau'}([e_j],[e_j])$ fails to be square summable, except for
$\tau>\tau'$ in a set of measure zero . Recall that 
\[
W_{SJ,\tau}(t,\ux;t',\ux') = \sum_{j\in J} 
\frac{\|S_j\|}{2\omega_j \|C_j\|}
\left(
e^{-i\omega_j t} + i \delta_j \sin \omega_j t \right)
 \left(
e^{i\omega_j t'} - i \delta_j \sin \omega_j t'\right) \psi_j(\ux)\overline{\psi_j(\ux')}\,,
\]
where $S_j$, $C_j$ and $\delta_j$ are defined with respect to $\tau$. We denote by
$S_j'$, $C_j'$ and $\delta_j'$ the corresponding values defined with respect to $\tau'$. Then
notice that
\[
 W_{SJ,\tau}(f,h) - W_{SJ,\tau'}(f.h) = {:} W_{SJ,\tau} {:}(f,h) - {:} W_{SJ,\tau'} {:}(f,h)\,, \quad
 f,h \in S(\Mb_{\tau'})\,.
\]
We construct the required orthonormal system as follows. As in the previous section, we can
assume that the basis $\psi_j = \psi_{({\rm j},\gamma)}$ of eigenvectors of $K$ for eigenvalues
$\omega_j$ is formed by real-valued, smooth functions on $\Sigma$. Then we choose for each
${\rm j} \in \mathbb{N}_0$ some real-valued function $\eta_{\rm j} \in C_0^\infty(-\tau',\tau')$ which is 
symmetric, $\eta_{\rm j}(-t) = \eta_{\rm j}(t)$, and normalized such that
\begin{align}
\frac{\|S_j\|}{2\omega_j \|C_j\|}\left( \int_{-\tau'}^{\tau'} \eta_{\rm j}(t) \cos(\omega_j t)\,dt\right)^2  & = 
 1 \,.
\end{align}
With these choices, we set
$$
e_j = e_{({\rm j},\gamma)} = \eta_{\rm j} \otimes \psi_{{\rm j},\gamma} \,.
$$
Then $e_j$ is smooth, compactly supported and real-valued, and one obtains that 
\[
 \mu_{SJ,\tau}([e_j],[e_{j'}]) = W_{SJ,\tau}(e_j,e_{j'}) = \delta_{j j'}\,,
\]
where the symbol on the right hand side is Kronecker's $\delta$. 
On the other hand, inserting the $e_j$ yields for $\mu_{SJ\tau}([e_j],[e_j]) - \mu_{SJ\tau'}([e_j],[e_j])$ the 
expression
\[
 {:} W_{SJ,\tau} {:}(e_j,e_j) - {:} W_{SJ,\tau'} {:}(e_j,e_j)
 = \frac{\|C_j\|}{\|S_j\|} \left( \frac{\delta_j}{(1 - \delta_j)} - \frac{\delta'_j}{(1 - \delta'_j)} \right)
= \frac{\|C_j\|(\delta_j - \delta_j')}{\|S_j\|(1 -\delta_j)(1 - \delta'_j)}
\,.
\]
The sequence $\|C_j\| / \|S_j\|$ converges to 1 as
${\rm j} \to \infty$, so we left with having to decide on summability of
$(\delta_j - \delta_j')^2$. Exploiting again \eqref{delta-asympt}, summability 
of $(\delta_j - \delta_j')^2$ requires
convergence of the sum
\[
 \sum_{{\rm j},\gamma}
  \left( \frac{\frac{\tau'}{\tau}\sin(2\omega_{\rm j}\tau) - \sin(2\omega_{\rm j}\tau')}
   {2\omega_{\rm j}\tau'} \right)^2\,,
\]
for which it is necessary that
\begin{align} \label{sintaulimit}
 \frac{\tau'}{\tau}\sin(2\omega_{\rm j}\tau) - \sin(2\omega_{\rm j}\tau') \to 0 \quad \text{as} \ \ 
 {\rm j} \to \infty\,.
\end{align} 
Now we argue that  this holds at most for a set of values of pairs $\tau > \tau'$
which has measure equal to zero. Let us fix some $\tau_0 > \tau_0' > 0$. We want to show that the
set $\widetilde{V}$,  formed by all pairs $(\tau',\tau) \in \mathbb{R}^2$ with 
$\tau_0 > \tau > \tau' > \tau'_0$ 
and having the
property that
\eqref{sintaulimit} holds, has zero Lebesgue measure.
Obviously it is sufficient to show that the set $\widetilde{Q}$, 
formed by all pairs $(\varrho,\tau)$ with
$\tau_0 > \tau > \tau_0'\,,\ 1 > \varrho > \varrho_0 $ for suitable $\varrho_0 > 0$,
and fulfilling
\[
\varrho^2 \sin^2(2\omega_{\rm j}\tau) - \sin^2(2\omega_{\rm j}\tau\varrho) \to 0 \quad ({\rm j} 
\to \infty)\,,
\]
has zero Lebesgue measure (it is easily seen to be a Borel set).
Using trigonometric identities and applying the theorem of
dominated convergence, the definition of $\widetilde{Q}$ implies
\begin{align} \label{cosconvergence}
\int_{\widetilde{Q}} (1 - \varrho^2) \,d\varrho\,d\tau -
\int_{\widetilde{Q}} (\cos(4\omega_{\rm j}\tau \varrho) - \varrho^2\cos(4\omega_{\rm j} \tau))\,d\varrho\,d\tau
\to 0 \quad ({\rm j} \to \infty)\,.
\end{align}
However, if $I_1$ and $I_2$ are any open, relatively compact intervals contained in $(0,\infty)$,
the Riemann-Lebesgue Lemma entails that
\begin{align} \label{Rie-L-conver}
\int_{I_1 \times I_2} (\cos(4\omega_{\rm j}\tau \varrho) - \varrho^2\cos(4\omega_{\rm j} \tau))\,d\varrho\,d\tau
\to 0 \quad ({\rm j} \to \infty)\,.
\end{align}
As the
set of step-functions is dense in the $L^2$-space of any bounded, measurable
subset of $\mathbb{R}^2$, and as 
the $\cos(...)$ functions are uniformly bounded in ${\rm j}$, one concludes that 
\eqref{Rie-L-conver}
generalizes
to the case where $I_1 \times I_2$ is replaced by $\widetilde{Q}$. 
Thus, \eqref{cosconvergence} 
implies $\int_{\widetilde{Q}}(1 -\varrho^2)\,d\varrho\,d\tau = 0$, and this can only be fulfilled if 
$\widetilde{Q}$ has zero Lebesgue measure, concluding the proof.
${}$ \hfill $\Box$

Let us briefly comment why the result of Proposition \ref{disaster} 
virtually excludes the class of S-J states as a class of physical
states. While the result on disjointness of S-J from Hadamard states is
bad enough in the light of the vast amount of results in support of
the idea that Hadamard states are, indeed, the appropriate generalization of
vacuum-like or thermal-state-like states to quantum field theory in curved spacetime
--- and thereby the best candidates for physically realizable states --- one could take
the point of view that another class of states were more suitable. Nevertheless,
for any specification of a class of physical states, one would require that the
states of that class are (locally) quasiequivalent. Otherwise, as mentioned before,
taking different states in the class as reference states might give rise to entirely
different notions of particle numbers --- typically, if two states are disjoint, the first
state cannot be realized as a state of finite particle number in the GNS representation
of the second state, and vice versa. For S-J states defined on ultrastatic slab
spacetimes with different time-intervals, Proposition \ref{disaster} shows that this
most unfavourable situation prevails. In particular, there is no way to base a 
particle interpretation, 
and estimates related to cosmological particle production, on using S-J states
as reference states,
as proposed by~\cite{AAS}.   
 
\section{Local covariance}\label{sect:loccov}

As noted in~\cite{AAS}, the S-J state depends in a nonlocal way on the spacetime region used in its construction. Consider nested globally hyperbolic spacetime regions $V\subset W$ on which the S-J prescription is well-defined, yielding states $\omega_{SJ}^{(V)}$
and $\omega_{SJ}^{(W)}$ respectively.  Either state can be used to provide expectation values for observables localised in $V$, such as products of fields smeared with 
test functions supported in $V$, and these values may differ. This may be seen 
explicitly in the example of the ultrastatic slab, because of the dependence of
$\|S_j\|$, $\|C_j\|$ and $\delta_j$ on the value of $\tau$. Thus, for example,
if  $V=(-\tau_V,\tau_V)\times\Sigma$
and $W = (-\tau_W,\tau_W)\times\Sigma$, with $0<\tau_V<\tau_W$,
then the resulting S-J two-point functions differ on test functions
supported in $V$ -- this is easily seen for test functions of form $f(t)\psi_j(\ux)$,
with $f\in\CoinX{-\tau_V,\tau_V}$.

This raises problems of intepretation. In order to compute the expectation value
of a given local observable using the S-J prescription, it is necessary to decide on a 
choice of spacetime region to compute a suitable S-J state, and the expectation
value obtained depends on this choice. 

Accordingly the S-J prescription does not produce unique expectation values for 
local physical quantities unless a way can be found to fix the choice of region.
One way of doing this might be to declare that the region in will always be
chosen to be the whole spacetime (setting aside questions of whether 
the prescription is well-defined on unbounded regions). However, the ambiguity
is not removed in cases where the spacetime can be isometrically embedded
as a causally convex subspacetime of a larger one, leading once more to
competing prescriptions on the original spacetime.

On these grounds, it seems to us that the S-J prescription has rather limited
operational significance.\footnote{That would not necessarily prevent it
being of technical use, were it (or a variant thereof) Hadamard.}
Turning this around, an operationally sound prescription for a single distinguished
state should be such that the prescription on region $W$ restricts to any
suitable subspacetime $V\subset W$ to give the state directly assigned to $V$ by
the prescription. This can be made precise in the context of the functorial formulation
of quantum field theory in curved spacetimes introduced in~\cite{BrFrVe03}. In
\cite{FewVer:dynloc_theory},  we defined a {\em natural state} to be precisely a choice
of state $\omega_\Mb$ to each globally hyperbolic spacetime $\Mb$, which respects spacetime embeddings:\footnote{Embeddings
are required to be isometric, preserve (time-)orientation and to have causally
convex image.} specifically, each such embedding $\psi$ of globally hyperbolic spacetime $\Mb$ in globally hyperbolic spacetime $\Nb$ induces a $*$-homomorphism
of the corresponding $*$-algebras $\Af(\psi):\Af(\Mb)\to\Af(\Nb)$, and we
demand that the distinguished states [which, in particular, belong to the dual spaces of the algebras] should obey $\omega_\Mb = \Af(\psi)^*\omega_\Nb$.

The aim is to prove that such choices do not exist. To this end, we proved
a no-go theorem that can be paraphrased as follows: if the theory, when
formulated in Minkowski space, obeys standard assumptions of Wightman or
Haag--Kastler QFT, with the distinguished state as the Minkowski vacuum, and
the theory in general spacetimes obeys the conditions of extended locality and dynamical locality, then the theory is trivial -- its algebra consists, in any spacetime, only of complex multiples of the identity. Extended locality is simply the requirement that
local algebras of spacelike separated regions should intersect only in multiples of
the identity; dynamical locality was introduced in~\cite{FewVer:dynloc_theory} -- while physically motivated, it requires a careful definition that would take us too far from the present subject.  We refer to \cite{FewVer:dynloc_theory} for the details, and \cite{FewsterRegensburg} for a short
overview. Note, however, that several standard models are known to satisfy dynamical locality, including the massive minimally coupled scalar field~\cite{FewVer:dynloc2}, the nonminimally coupled scalar field
of any mass $m\ge 0$ and the algebra of Wick polynomials of scalar
fields (at least in the cases of minimal and conformal coupling, in the former case
requiring nonzero mass)~\cite{Ferguson:2012}. The massless minimally coupled scalar
field is a special case, owing to its rigid gauge invariance $\phi\mapsto\phi+\text{const}$; once this is taken into account, dynamical locality is restored~\cite{FewVer:dynloc2}.

We also emphasise that our no-go result is not restricted to free-field models, but
applies to any theory obeying our conditions, and also that the version
stated here is rather weaker than the full statement; for instance, it also 
applies even if the distinguished state does not coincide with the Minkowski
vacuum state, provided there is some spacetime with noncompact Cauchy surfaces in which it induces a representation with the Reeh--Schlieder property.

Summarising, the S-J prescription evades our no-go result because it is not
locally covariant; however, this severely limits its claim to be a distinguished state,
from an operational point of view.

\section{Conclusion}

We have presented a number of problems with the S-J state prescription. 
First, on ultrastatic slab spacetimes $(-\tau,\tau)\times\Sigma$ with compact
spatial section $\Sigma$, the S-J state was shown not to be Hadamard except, at most,
for a set of $\tau$ with measure zero (and in two concrete examples, there were
no exceptional values). This already essentially rules out the S-J states from
physical interest if one accepts the standard theory of renormalisation of the
stress-energy tensor and other Wick products using point-splitting. Second, we showed
that the representation induced by S-J state is disjoint from (in particular, unitarily inequivalent to)
both the Hadamard representation induced by the restriction of the ultrastatic ground state to the slab, and 
the representation induced by an S-J state corresponding to a different value of $\tau$. Again, there
is the possibility of exceptional values of the parameters, with at most measure zero.
Although the results on disjointness were proved for the case where the spatial section is
the $3$-sphere, we see no problem in principle with extending these results to more general
ultrastatic slabs. Thus, even if one rejects the point-splitting ideology, it would appear that
the S-J states do not offer a good alternative notion of particles. Third, we have discussed
the wider interpretational problems arising from the lack of local covariance in the S-J prescription. 

Along the way, we have proved some positive statements about the S-J proposal, which may be
useful in any future development. We emphasise that we have not excluded the possibility that
S-J states might be Hadamard in relatively compact regions with nontrivial causal complement,
although this seems unlikely to us. It is also conceivable that a modification of the prescription
might lead to Hadamard states -- it will be clear from our discussion that most of the problems
arise because the coefficients appearing in expressions such as Eq.~\eqref{eq:WS_normord}
decay rather slowly in $j$. Perhaps a modified prescription might resolve such problems and
be of technical utility, although we emphasise that the no-go result of~\cite{FewVer:dynloc_theory} rules out any locally covariant prescription and therefore limits the physical significance of any resulting
states. 

Finally, we mention for completeness that Hadamard states can be constructed in (classes of) general spacetimes in various ways beyond the `deformation argument' of~\cite{FullingNarcowichWald},
for example using the BMS group in asymptotically flat spacetimes~\cite{Moretti:2008}. 
In the context of cosmology on Robertson--Walker spacetimes, specific Hamadard
states have been considered more recently. This includes locally thermal states 
\cite{VerchRegensburg} and states with distinguished symmetry and thermal properties
at the conformal spacetime boundary \cite{DapHackPin-RW-KMS:2011}. Another class
are states of minimal (local) energy (which in a way seem to come close to
a modification of S-J states which renders Hadamard states) introduced by Olbermann
\cite{Olbermann:2007}; cosmological particle creation can be calculated rigorously
for this class of states \cite{DegnerVerch:2010}. Finally,
 a detailed study and construction of the class of adiabatic states can be
 found in~\cite{JunkerSchrohe:2002}; see also \cite{NN1,NN2} for some earlier references.
\\[10pt]
{\noindent\em Acknowledgments} CJF thanks SP Eveson for useful conversations concerning detailed points of measure theory.

\appendix
\section{Proof of Proposition~\ref{prop:bdedA}}

Suppose $\Mb$ and $\Nb$ are globally hyperbolic spacetimes and that there is an isometric embedding $\psi:\Mb\to\Nb$, preserving orientation and time-orientation, and so that $\psi(M)$ is a causally convex and relatively compact subset of $N$. We will sketch the proof that the advanced-minus-retarded distribution on $\Mb$ defines a bounded operator on $L^2(\Mb)$. 

By uniqueness of the advanced and retarded fundamental solutions and the properties of $\psi$,
it is easily shown that $\Ett^\pm_\Mb f = \psi^* \Ett^\pm_\Nb \psi_* f$ for all $f\in\CoinX{\Mb}$, where
$\psi^*$ is the pull-back and $\psi_*$ is the push-forward,
\[
(\psi_* f)(x) = \begin{cases} f(\psi^{-1}(x)) & x\in \psi(M) \\ 0 & \text{otherwise}.\end{cases}
\]
Hence we also have $\Ett_\Mb f = \psi^* \Ett_\Nb \psi_* f$ for such $f$. Now as $\Ett_\Nb \psi_* f\in C^\infty(N)$ and $\psi(M)$ is relatively compact, $\Ett_\Mb f$ must be bounded and therefore (as
$\Mb$ has finite volume) we have $\Ett_\Mb f\in C^\infty(M)\cap L^2(\Mb)$ for all $f\in\CoinX{M}$. 

As $\Nb$ is globally hyperbolic, it may be identitied with $\RR\times\Sigma$ so that each 
$\{t\}\times\Sigma$ is a Cauchy surface, and with the metric splitting as $ds^2 = \beta(t,\ux)dt^2 - 
h_{ij}(t,\ux) dx^i dx^j$, where $x^i$ are local coordinates on $\Sigma$. With respect
to such a foliation, a standard energy estimate gives the following result (see, e.g., 
\cite{HawkingEllis} for similar results): {\em 
For any compact interval $I\subset \RR$ and relatively compact open $O\subset I\times \Sigma$,
there is a constant $C_{I,O}>0$ such that
\[
\|\Ett^\pm_\Nb f\|_{L^2(I\times\Sigma)} \le C_{I,O} \| f\|_{L^2(I\times\Sigma)}
\]
for all $f\in\CoinX{O}$, where the volume measure on $\Nb$ is used to define the $L^2$ norms.} 
As $\psi(M)$ is relatively compact, we may choose $I$ so that $\psi(M)\subset I\times\Sigma$, 
whereupon we find
\[
\|\Ett^\pm_\Mb f\|_{L^2(\Mb)} \le \|\Ett^\pm_\Nb\psi_* f\|_{L^2(I\times\Sigma)} 
 \le C_{I,\psi(M)} \|\psi_*  f\|_{L^2(I\times\Sigma)} = C_{I,\psi(M)} \| f\|_{L^2(\Mb)}.
\]
and hence we may conclude that $\Ett_\Mb$ defines a bounded operator on $L^2(\Mb)$
with operator norm less than $2C_{I,\psi(M)}$.

\vspace{-0.5\baselineskip}

\end{document}